\documentclass[12pt,preprint]{aastex}
\usepackage{graphicx,epsfig,epsf,amssymb}      

\newcommand{\myemail}{litp@ihep.ac.cn}

\def \<{\langle}
\def \>{\rangle}

\def \tr{$\sqrt{\<\hat{t}^2\>}$}
\def \t{$\mathbf{\hat{t}}$}
\newcommand{\degree}{^\circ}

\shorttitle{Errors in CMB maps}
\shortauthors{Liu \& Li}

\begin{document}
\title{ Statistical and systematical errors\\ in cosmic microwave background maps}
\author{Hao Liu\altaffilmark{1,3} and Ti-Pei Li\altaffilmark{1,2}}
 \altaffiltext{1}{Key Lab. of Particle Astrophys., Inst. of High Energy Phys.,
Chinese Academy of Sciences, Beijing}
\altaffiltext{2}{Center for
Astrophysics, Tsinghua University, Beijing, China; \myemail}
\altaffiltext{3}{Graduate School of Chinese Academy of Sciences, Beijing}

\begin{abstract}
Sky temperature map of the cosmic microwave 
background (CMB) is one of the premier probes of cosmology.  
To minimize instrumentally induced systematic errors, 
CMB anisotropy experiments measure temperature differences across 
the sky using paires of horn antennas with a fixed separation angle, 
temperature maps are recovered from temperature
differences obtained in sky survey through a map-making procedure.
 The instrument noise, inhomogeneities of the sky coverage and sky temperature 
inevitably produce statistical and systematical errors in recovered 
temperature maps.
We show in this paper that observation-dependent noise and systematic temperature
distortion contained in released Wilkinson Microwave Anisotropy Probe (WMAP) 
CMB maps are remarkable. These errors 
can contribute to large-scale anomalies detected 
in WMAP maps and distort the angular power spectrum as well.
It is needed to remake temperature maps from original WMAP differential data
with modified map-making procedure 
to avoid observation-dependent noise and systematic
distortion in recovered maps. 
\end{abstract}


\section{Map-making}
The COBE and WMAP missions measure temperature differences between sky points 
using differential radiometers consisting of plus-horn and minus-horn~\cite{smo90,ben03a}. 
Let denote $t_i$ the temperature anisotropy at a sky pixel $i$. 
The raw data in a certain band is a set of 
temperature differences {\bf d} between pixels in the sky. From $N$ observations we have 
the following observation equations 
\begin{equation}
\label{dt}
\begin{array}{c@{\:-\:}c@{\;=\;}c}
t_{1^+} & t_{1^-} & d_1 \\
t_{2^+} & t_{2^-} & d_2 \\
\multicolumn{3}{c}{\dotfill}\\
t_{N^+} & t_{N^-} &~~d_N~. 
\end{array}
\end{equation}                                                                     
The above equation system can be expressed by matrix notation
\begin{equation}
\label{dt1}
\mathbf{At=d}~.
\end{equation}
Where the scan matrix of the experiment {\bf A}$ =(a(k,i)),~k=1,\cdots,N$ 
and $i=1,\cdots,L$ with $L$ being the total number 
of sky map pixels. The most of elements $a(k,i)=0$ except for $a(k,i=k^+)=1$   
and $a(k,i=k^-)=-1$, where $k^+$ denotes the pixel observed by the plus-horn 
and $k^-$ the pixel observed by the minus-horn at an observation $k$.

The normal equation of Eq.~\ref{dt} or Eq.~\ref{dt1} is
\begin{equation}
\label{ne}
\mathbf{Mt=A^Td}
\end{equation}
with $\mathbf{M=A^TA}$.
 The least-squares estimate 
of the sky map results from solving Eq.~\ref{ne}
\[ \mathbf{\hat{t}=M^{-1}A^Td}~. \]
The WMAP team \cite{hin03} uses the following approximate formula to compute the 
iterative solution 
\begin{equation}
\label{mm-w}
\mathbf{t^{(n+1)}=\tilde{M}^{-1}(A^Td-A^TAt^{(n)})+t^{(n)}}~,
\end{equation}
where $\mathbf{\tilde{M}^{-1}}=$ diag$(\frac{1}{N_1},\frac{1}{N_2},\cdots)$  
is an approximate inverse of {\bf M} with $N_i$ being the total number
of observations for pixel $i$. 

The use of approximate inverse matrix  $\mathbf{M^{-1}}$ is not necessary.
Here we derive an iterative formula directly from the normal equation.  
The Eq.~\ref{ne} can be expressed as
\begin{eqnarray*}
N_i^+t_i-\sum_{k^+=i}t_{k^-}-\sum_{k^-=i}t_{k^+}+N_i^-t_i
=\sum_{k^+=i}d_k-\sum_{k^-=i}d_k~,  \\ 
 (i=1,2,\cdots,L)~.
\end{eqnarray*}
Where $\sum_{k^+=i}$ means summing over $N_i^+$ observations while the pixel 
$i$ is observed by the plus-horn and $\sum_{k^-=i}$ means summing over $N_i^-$
observations while the pixel $i$ is observed by the minus-horn, 
and the total number of observations for the pixel $i$ is $N_i=N_i^++N_i^-$. 
From the above equations we can derive the following iterative formula
\begin{eqnarray}
\label{mm-l}
t_i^{(n+1)}=\frac{1}{N_i}(\sum_{k^+=i}(d_k+t_{k^-}^{(n)})-\sum_{k^-=i}
(d_k-t_{k^+}^{(n)}))~, \nonumber\\
(i=1,2,\cdots,L)~.
\end{eqnarray}

With Eq.~\ref{mm-w} or Eq.~\ref{mm-l} when the number $n$ of iteration is large enough, we get 
the final solution $\hat{t}_i=t_i^{(n)}$ for each pixel $i$.
The Eg.~\ref{mm-w} used by the WMAP team is an approximate formula and Eq.~\ref{mm-l} 
is an exact one, but both has good performance for the differential data of a 
noiseless instrument. With Eq.~\ref{mm-l} we can easily study the statistical 
and systematical errors
induced by instrument noise, inhomogeneity of sky coverage, inhomogeneity of  
sky temperature, and unbalance between two sky side measurements.

\section{Exposure-dependent noise}
 We can directly see from Eq.~\ref{mm-l} that their exists exposure dependent noise in
a recovered temperature map.
The WMAP differencing assembly has instrument noise {\bf n} per observation, 
the real observation data $d_k=t_{k^+}-t_{k^-}+n_k$. The instrument noise is not negligible, 
e.g. the mean rms noise $\sigma_0=6.7$ mK for the W4-band of WMAP \cite{lim03}.  
The obtained temperature differences $d_k$ can be 
taken as a random variable with a standard error $\sigma_0$. 
The final iterative solution $\hat{t}_i$ from Eq.~\ref{mm-l} has a noise component 
being the mean of $N_i$ variables, i.e. the temperature ${\hat t}_i$ in a WMAP 
map for a sky pixel $i$  has an exposure-dependent error 
\begin{equation}
\label{sigma}
\sigma_i=\frac{1}{\sqrt{N_i}}\sigma_0~. 
\end{equation}

By analyzing CMB maps from the first year WMAP (WMAP1) data, 
Tegmark et al. (2003) find both the CMB quadrupole and octopole having
 power along a particular spatial axis and more works \cite{cos04,eri04a,sch04,jaf05} 
find that the axis of maximum asymmetry tends to lie close to the ecliptic axis. 
A similar anomaly was also found in COBE maps \cite{cop06}. 
The unexplained orientation of large-scale 
patterns of CMB maps in respect to the ecliptic frame is one of the biggest surprises 
in CMB studies \cite{sta05}.
A notable asymmetry of temperature fluctuation power in two opposing hemispheres is also found 
in the WMAP1 and COBE results  \cite{eri04b, han04}.
After the release of WMAP results in 2006 March, similar large-scale anomalies 
are still detected 
in the WMAP3 data \cite{abr06,jaf06,cop07,lan07,eri07,par07,vie07,sam08}.       
The unexpected large-scale anomalies in CMB maps are extensively studied  
with different techniques, but their reasons still remain unclear.
Here we show that the exposure-dependent noise should be an important 
source of detected anomalies. 

\subsection{Large-scale non-Gaussian modulation}

 \begin{figure}[p]
    \label{f1}
   \vspace{2mm}
   \begin{center}
   \psfig{figure=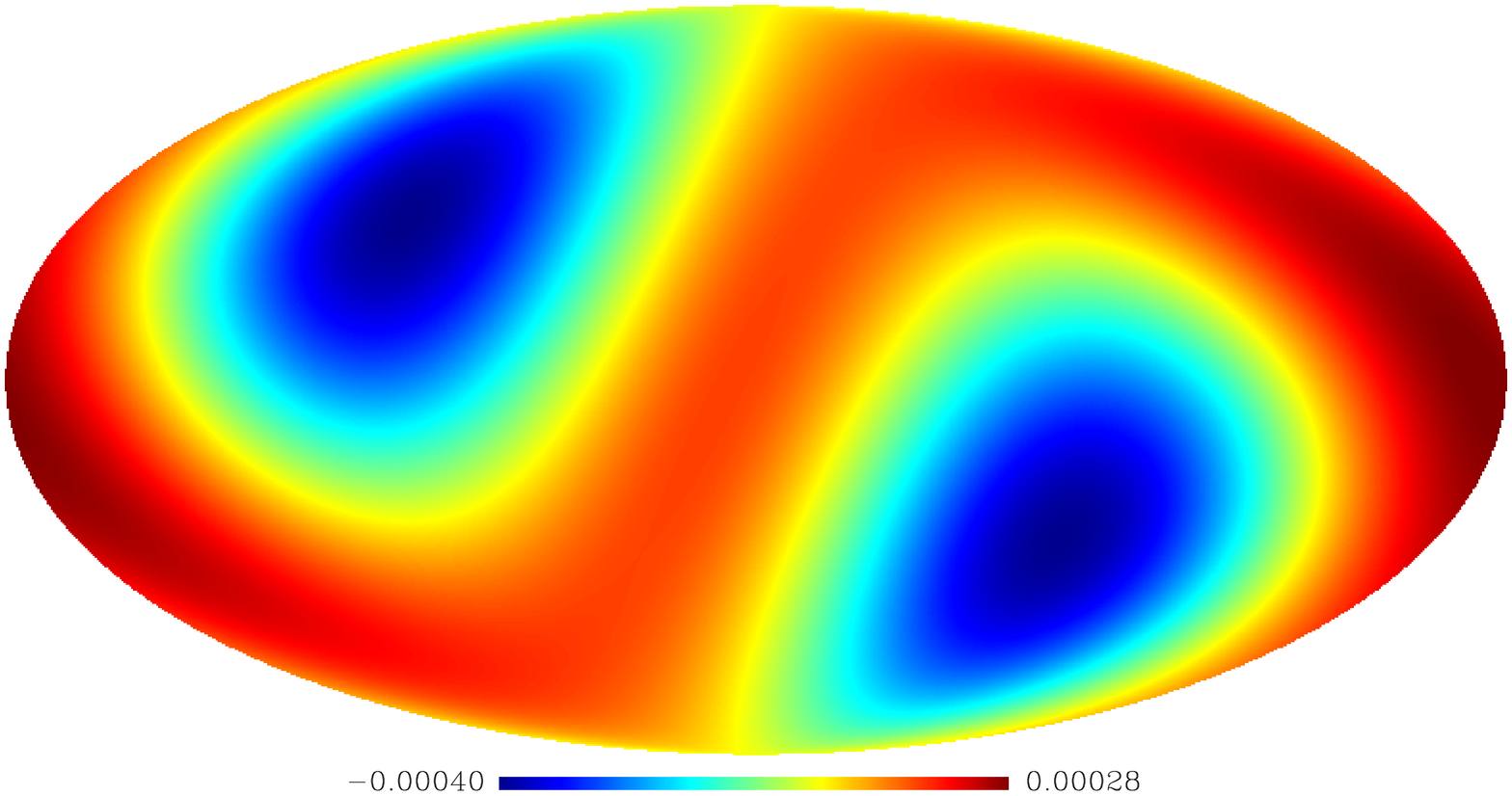,width=45mm,height=70mm,angle=90}\\
\vspace{-5mm}\psfig{figure=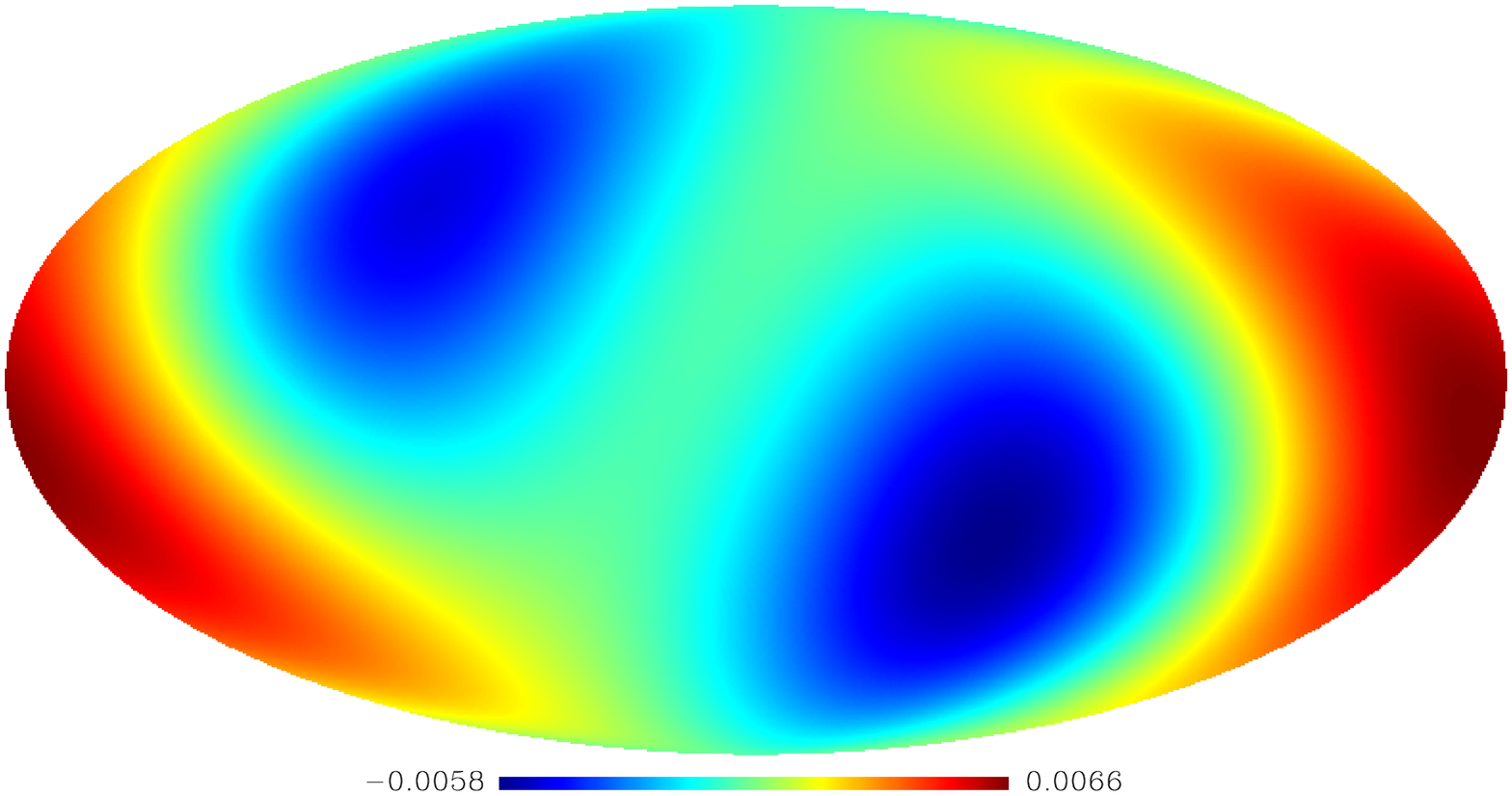,width=45mm,height=70mm,angle=90}\\
\vspace{-5mm}\psfig{figure=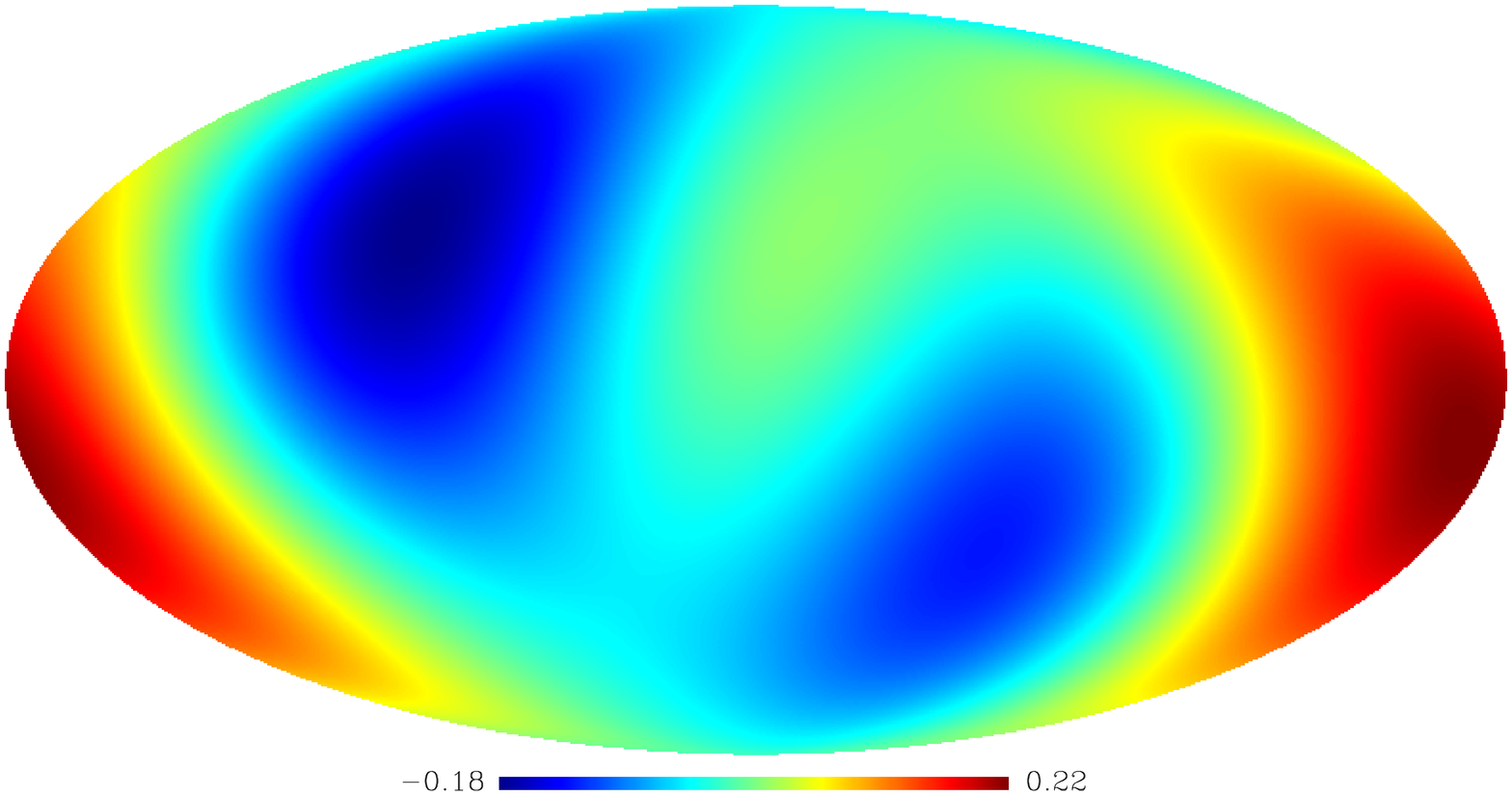,width=45mm,height=70mm,angle=90}
 \end{center}
\caption{Large-scale non-Gaussian modulation features with $l_{max}=2$
for the WMAP3 V-band data . 
{\sl Top panel}: large-scale feature of $1/N$ map.  
{\sl Middle panel}: large-scale feature of $1/\sigma(N)$ map. 
{\sl Bottom panel}: Best-fit large-scale modulation function $f(\mathbf{\hat{n}})$
for temperature map.}     
\end{figure}

Because the sky coverage of WMAP mission is inhomogeneous
-- the number of observations being greatest at the ecliptic poles 
and the ecliptic plane being most sparsely observed~\cite{hin07}, from 
Eq.~\ref{sigma} we know that the WMAP temperature maps 
contain inhomogeneous exposure-dependent noise components.
The remarkable fact, the feature of large-scale anomalies 
detected in WMAP maps being very similar with the WMAP exposure pattern,
strongly indicates the existence of such observational effect on WMAP maps.  
To address the large scale anomalies, such as asymmetry, 
alignment and low $l$ power issues detected in WMAP data
with different techniques, the WMAP team~\cite{sper06} describe the observed 
temperature fluctuations, $\mathbf{{\hat{t}}}$, as a Gaussian and isotropic random 
field,  $\mathbf{t}$, modulated by a function $f(\mathbf{n})$
\[ \hat{t}(\mathbf{n})=t(\mathbf{n})[1+f(\mathbf{n})] \]
where $f(\mathbf{n})$ is an arbitrary modulation function. They expand $f(\mathbf{n})$
in spherical harmonics 
\[ f(\mathbf{n})=\sum_{l=1}^{l_{max}}\sum_{l=-m}^mf_{lm}Y_{lm}(\mathbf{n}) \]
and use maximum likelihood technique with 
a Markov Chain Monte Carlo solver to get the best fit values of $f_{lm}$ 
with $l_{max}=2$ for the WMAP3 V-band map.  
 The bottom panel of Fig.~1 is obtained
based on the best fit coefficients, showing
in a unifying manner the large scale anomalies in WMAP temperature fluctuations
which is the same feature that has been identified in a number 
of papers on non-Gaussianity.
We calculate the spherical harmonic coefficients $f_{lm}$ with $l_{max}=2$ 
for the map of $1/N$ with $N$ being number of observations per sky pixel 
from the WMAP3 V-band data. The top panel of Fig.~1 shows 
the map of $1/N$ reconstructed based on the coefficients $f_{lm}$.  
The middle panel shows the reconstructed result for
the observation fluctuation map --
the map of $1/\sigma(N)$ where the rms variation $\sigma(N)=\sqrt{\<(N-\<N\>)^2\>}$
calculated within a region of $\sim 1\degree$ side dimension for each sky pixel.
The large-scale non-Gaussian modulation features of WMAP temperature map 
and scan pattern being similar for each other suggests that large scale anomalies 
detected in WMAP maps are most probably resulted from observation effect, not
cosmological origin. In comparing the top and middle panels in  Fig.~1, 
the modulation pattern for the observation fluctuation map  
is more similar to the detected anomalies shown in the bottom panel,
indicating that the fluctuation of observation numbers could produce additional
noise component to the recovered temperature map. 
        
\subsection{Alignment and planarity}
\begin{figure}[t]
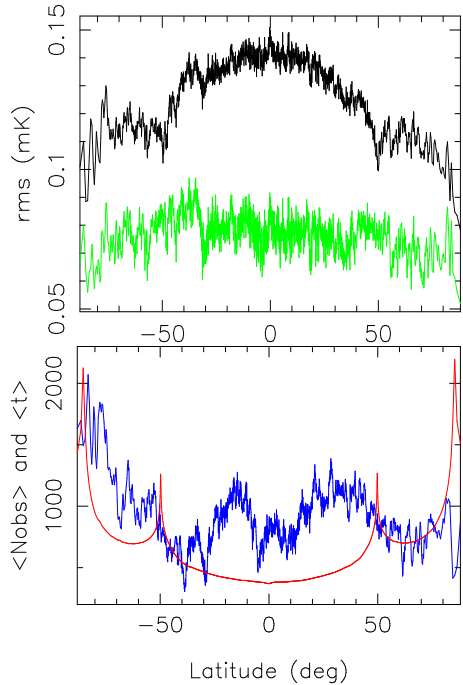

\label{f2}
   \begin{center}
   \psfig{figure=f2a.ps,width=45mm,height=60mm,angle=270}\\
\psfig{figure=f2b.ps,width=45mm,height=60mm,angle=270}
   \caption{Latitude distributions from the WMAP3 foreground-cleaned Q1-band 
map in ecliptic coordinates.
{\sl Black line}: rms variation in the map, $\sigma_0=2.27$ mK.
{\sl Green line}: residual rms after excluding the exposure-dependent anisotropy
component by Eq.~\ref{rms}.
{\sl Red line}: average number of observations per sky pixel.   
{\sl Blue line}: average map temperature (in mK, times by a factor of
$2.76\times 10^4$). 
}
   \end{center}
\end{figure}

To show the anisotropy noise, we compute the average rms variation $\sqrt{\<{\hat{t}}^2\>}$ 
for the WMAP3 foreground-cleaned Q1-band 
temperature map at different latitudes; the results are shown by the black 
line in Fig.~2. The red line in Fig.~2
shows that of average number of observations per sky pixel.   
To modify the exposure-induced rms anisotropy,
we use the following formula to get the residual variation 
\begin{equation}
\label{rms}
rms=\sqrt{\<({\hat{t}}-\<{\hat{t}}\>)^2\>-\<\sigma_0^2/N\> }~.
\end{equation}
The residual rms of WMAP3 Q1-band , the green line in Fig.~2, 
shows that the latitude dependences are well excluded
by using Eq.~\ref{rms}, or, in other words, the noise per pixel in
the WMAP map is really exposure-dependent in a way described by Eq.~\ref{sigma}.   

\begin{figure}[t]
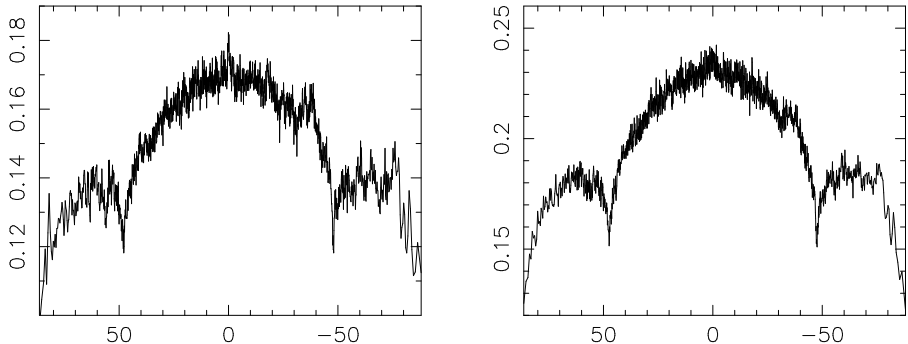

\label{f3}
   \begin{center}
    \psfig{figure=f3a.ps,width=45mm,height=55mm,angle=270}\hspace{8mm}
    \psfig{figure=f3b.ps,width=45mm,height=55mm,angle=270}
   \caption{Latitude distributions of temperature rms variation 
from WMAP3 foreground-cleaned maps in ecliptic coordinates.
{\sl left}: V1 band, $\sigma_0=3.29$ mK.
{\sl right}: W1 band, $\sigma_0=5.83$ mK.
}
   \end{center}
\end{figure}
\begin{figure}[t]
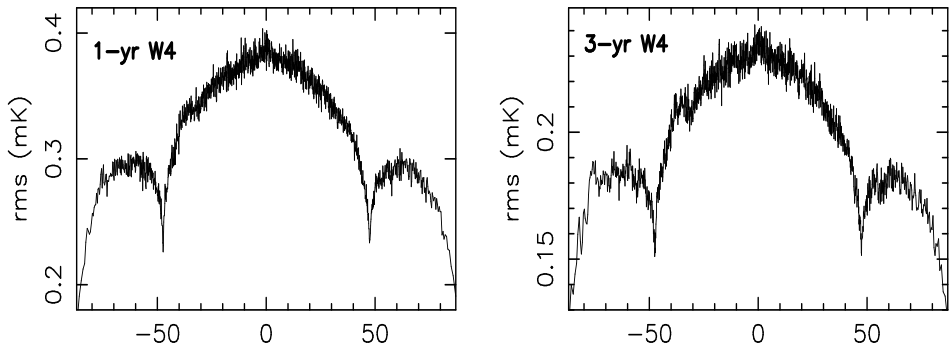

\label{f4}
   \begin{center}
    \psfig{figure=f4a.ps,width=45mm,height=60mm,angle=270}\hspace{4mm}
    \psfig{figure=f4b.ps,width=45mm,height=60mm,angle=270}
   \caption{Latitude distributions of temperature rms variation 
from WMAP foreground-cleaned maps in ecliptic coordinates.
{\sl left}: WMAP1 W4 band, $\sigma_0=6.7$ mK.
{\sl right}: WMAP3 W4 band, $\sigma_0=6.7$ mK.
}
   \end{center}
\end{figure}

Similar features are also found
in other waveband maps and the combined frequency maps (TOH and ILC maps) 
as well. 
For example, Fig.~3 shows the results from WMAP3 V1 and W1 bands. 
The similarity of structures of latitude distribution of rms variation for different  bands shown 
in Fig.~2 and Fig.~3 indicate that they are commonly originated 
from the ununiformity of sky exposure.      
The average pixel rms is 0.20 mK from $N=1.68\times 10^6$
observations for W1 band, 0.15 mK from $1.12\times 10^6$ observations for V1 band and 
0.13 mK from $0.84\times 10^6$ observations for Q1 band.
 The instrument noise $\sigma_0$ for W1, V1 and Q1 band are 5.85, 3.29 and 
2.27 mK respectively \cite{lim03}. We find that the ratios 
between rms$\times \sqrt{N}$ of different bands
are approximately equal to that between $\sigma_0$ of corresponding bands, 
which is what expected by Eq.~\ref{sigma}.  
Fig.~4 shows that the noise anisotropy for the three year WMAP W4-band map is 
almost the same as that for the first year map, not suppressed with
data accumulation.

As demonstrated above, the released WMAP temperature maps  
contain considerable exposure-dependent noise.
 The remarkable similarity 
between the large scale non-Gaussian modulation feature of WMAP temperature fluctuation
and exposure-pattern of WMAP observation, as shown in Fig.~1, suggests that 
large scale anomalies detected in WMAP maps are most probably resulted from 
observation effect, not astrophysical or cosmological origin.
 
For foreground removal the WMAP team produce the internal linear 
combination (ILC) map~${\mathbf t}$ from the five frequency sky maps 
${\mathbf t_i}~(i=1,\cdots,5)$ by
\(
 {\mathbf t}=\sum_iw_i{\mathbf t}_i \)
where the weights minimize the variance of final map, Var({\bf t}), under the constraint 
$\sum_iw_i=1$. 
For the region outside of the inner Galactic plane and by a nonlinear search, the weights $w$ 
are found to be 0.109, -0.684, -0.096, 1.921, -0.250 for
 K, Ka, Q, V, and W bands, respectively \cite{ben03c}. 
 Eriksen et al. (2004b) make minimization of the variance
under the constraint by means of Lagrange multipliers and obtain
the solutions are the inverse covariance weights
\(
 w_i=\frac{\sum_jC_{ij}^{-1}}{\sum_{j,l}C_{jl}^{-1}}
\)
with $C_{ij}$ being the map-to-map covariance matrix.

It is needed to inspect the effect of exposure dependent noise in the ILC map, 
which is extensively used 
in cosmological analysis, although the WMAP team warns against its use for 
CMB studies because of the complex noise properties of this map \cite{lim03}. 
The left panel of Fig.~5 shows the latitude distribution of rms deviation from
the WMAP3 ILC map. There exists a variation component monotonically decreasing from the 
southern pole to the northern pole, expressed roughly by the hand-drawn 
dotted line in the left panel of Fig.~5. The right panel shows the residual 
rms deviations after subtracting the north-south asymmetry component.
From Fig.~5 we see that the procedure of linear combination
with respect to the covariance between different frequencies can effectively  
depress the instrument noise. The latitude dependence of rms variation 
in the ILC map is weaker, but still visible with a structure similar 
to what observed in W, V and Q band. 
We can not make modification for the ILC map by sky exposure like what we do  
for Q, V and W data sets, because the ILC map is reconstructed by re-adding the data 
at different frequencies, it is difficult to assign an observation number $N_i$ to a pixel $i$.   
\begin{figure}[tb]
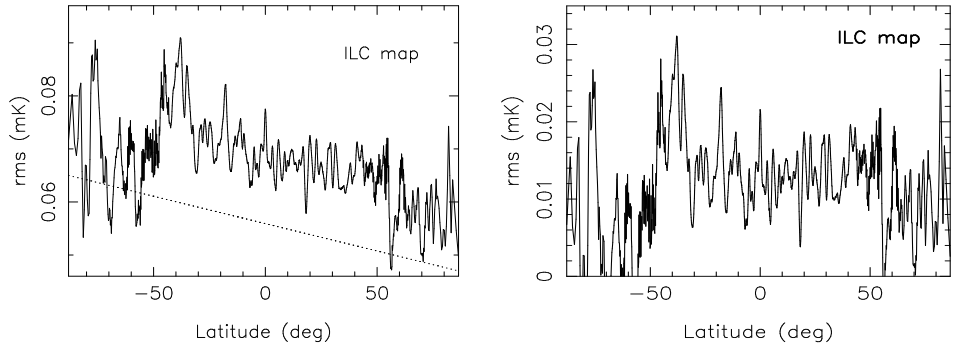

\label{f5}
   \begin{center}
    \psfig{figure=f5a.ps,width=45mm,height=60mm,angle=270}\hspace{4mm}
    \psfig{figure=f5b.ps,width=45mm,height=60mm,angle=270}
\end{center}
\caption{{\sl Left}: RMS deviation of temperatures vs. 
ecliptic latitude for the three-year ILC map. 
The vertical coordinate is \tr, where \t~denotes temperature anisotropies (in mK) 
at a certain latitude.
{\sl Right}: Residual rms deviations obtained from the curve shown in the left graph
 by subtracting the north-south asymmetry component 
(the dotted line in the left graph)} 
\end{figure}

\subsection{Pseudo sources}
\begin{figure}[tb]
\label{f6}
\vspace{-5.5cm}
\begin{center}
\includegraphics[width=14cm,angle=0]{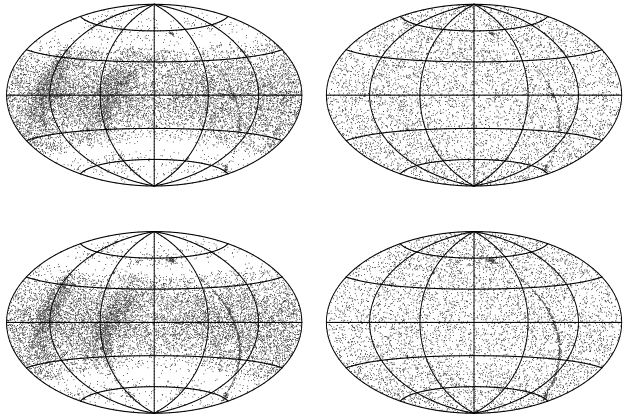}
\end{center}
\vspace{-9cm}
\caption{Cross-correlation map in ecliptic coordinates.  
Points show pixels with $c\ge 3\sigma_c$,
$\mathbf c$ -- cross-correlation of beam profile and sky temperature, 
$\sigma_c$ -- standard deviation of $\mathbf c$. {\it Left panel}, from W4-band map.  
{\it Right panel}, from modified W4-band map 
with multiplying temperatures by $\frac{\sqrt{M_i}}{\sqrt{\<M\>}}$.
{\it Top panel}, from WMAP1 data. {\it Bottom panel}, from WMAP3 data.} 
\end{figure}

In studying the nature of discrete $\gamma$-ray sources discovered by the COS-B satellite
experiment, we found that quite a lot apparent discrete sources 
revealed in the cross-correlation map with high significances (about half of sources in 2CG catalog) 
are pseudo-sources produced by the fluctuation of
structured diffuse background of the galactic plane \cite{lit82}. 
Here we use cross-correlation as another indicator to inspect
the possible large-scale anomaly  caused by the unevenly distributed noise in WMAP maps. 
We compute the cross-correlation function $\mathbf{c}$ of the 
W4-band map \t~and beam profile $\mathbf{B}$ by $\mathbf{c=B^T\hat{t}}$, 
and the standard deviation $\sigma_c$ of the correlation map.  The pixels with 
$|c|\ge 3\sigma_c$  in ecliptic 
coordinates are shown in the left panel of Fig.~6. That the apparent features of point-like 
sources concentrated along the ecliptic plane are just generated by observation effect 
as this feature disappears after modifying WMAP temperatures by a factor 
of $\frac{\sqrt{M_i}}{\sqrt{\<M\>}}$ for both one-year and three-year data, 
as shown by the right column of Fig.~6. The result shown by Fig.~6 remind us that 
one should take a care before to claim a finding of CMB anomaly from observation maps
before carefully considering the effect of exposure dependent noise.    
 
\subsection{North-south asymmetry}

\begin{figure}[htb]
\label{f7}
\begin{center}
\includegraphics[width=5cm,angle=270]{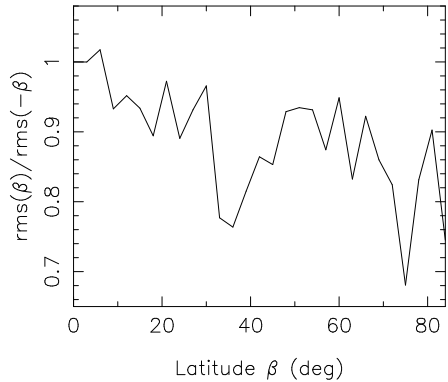}
\end{center}
\vspace{-0.1cm}
\caption{North-south asymmetry of rms deviation of temperatures in ILC map.
$\beta$ -- ecliptic latitude. 
rms($\beta$)/rms($-\beta$) -- ratio between rms deviations 
at the latitude $\beta$ in the northern hemisphere and at that in the southern hemisphere.}    
\end{figure}

\begin{figure}[htb]
\label{f8}
 \vspace{-8mm}
   \begin{center}
    \psfig{figure=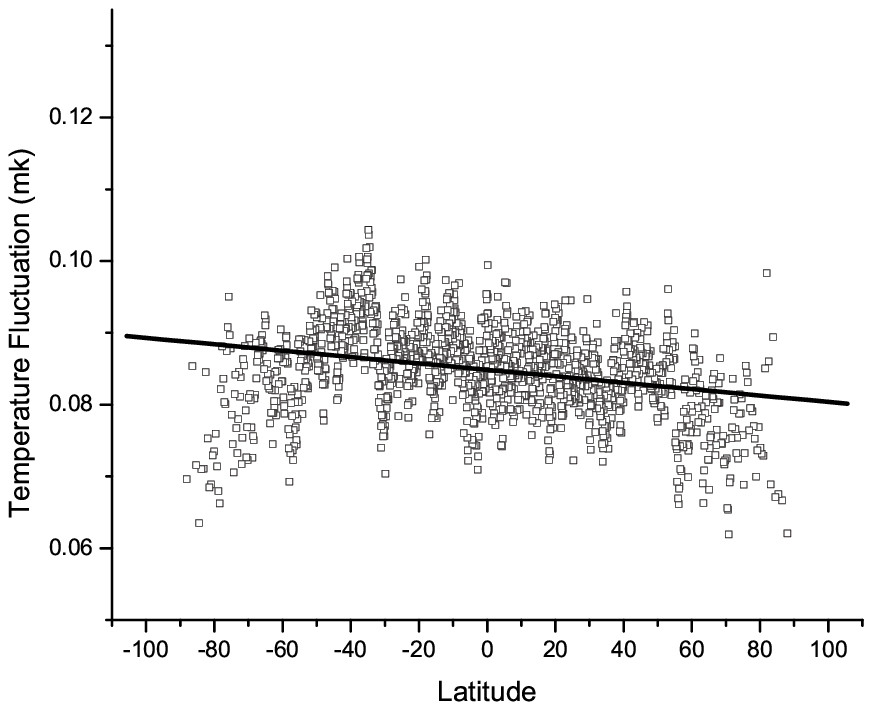,width=75mm,height=70mm,angle=0}
\vspace{-5mm}
   \caption{Ecliptic latitude dependence of residual temperature rms fluctuations 
of the foreground-cleaned WMAP3 V-band map.
The rms fluctuations are modified by using Eq.~\ref{rms}. The best-fit line
$y=-4.46\times10^{-5}x+0.085$.       
}
   \end{center}
\end{figure}

A remarkable character in Fig.~5 is
the monotonically decline trend of rms variation from the south pole to the north pole
in the ILC map. To show the north-south asymmetry quantitatively, we calculate the asymmetry ratio
rms$(\beta)$/rms$(-\beta)$ for different latitude $\beta$, where rms$(\beta)$ 
is the average rms of sky pixels with latitude within $\beta\pm1.5^o$ in the northern hemisphere 
and rms$(-\beta)$ is that in the south. The result is shown in Fig.~7. 
From Fig.~7 we see that there exists a 
systematic decrease of rms deviation from south to north with maximum asymmetry amplitude 
$\sim 10\%-20\%$. 
For different ecliptic latitudes we calculate  residual rms from the WMAP3 
 V-band foreground-cleaned
map after modifying the exposure ununiformity by using Eq.~\ref{rms},
the result is shown in Fig.~8, where we also find a clear north-south asymmetry.

\begin{figure}[tb]
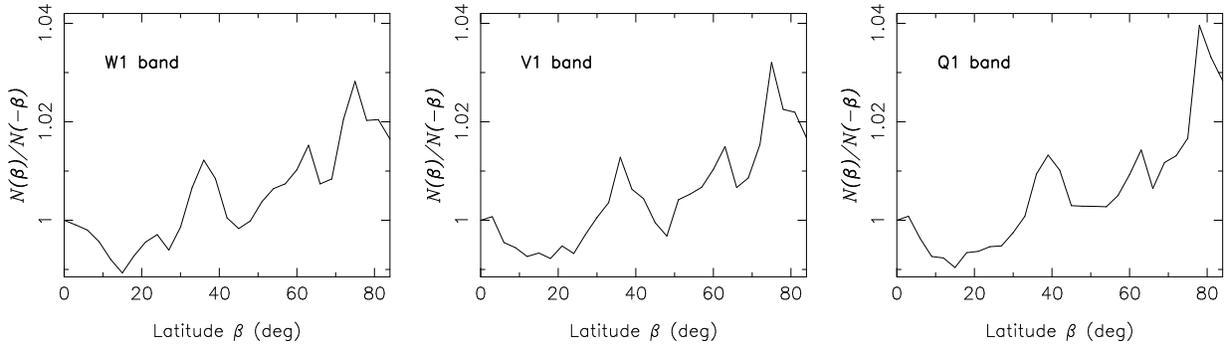

\label{f9}
\begin{center}
\vspace{-5mm}
 \includegraphics[width=4.5cm,angle=270]{f9a.ps}
 \hspace{3mm}\includegraphics[width=4.5cm,angle=270]{f9b.ps}
 \hspace{3mm}\includegraphics[width=4.5cm,angle=270]{f9c.ps}
\end{center}
\vspace{-0.1cm}
\caption{North-south asymmetry of WMAP3 exposure.   
 $N(\beta)$ is the total number of observations contributing to
the pixels with ecliptic latitude from $(\beta-1.5)$ deg to $(\beta+1.5)$ deg 
in the northern hemisphere,
 $N(-\beta)$ is that in the southern hemisphere.  
{\it Left panel}, W1 band. {\it Middle panel}, V1 band. {\it Right panel}, Q1 band.}
\end{figure}

To see if the north-south asymmetry also exists in WMAP exposure,
we calculate
the exposure asymmetry ratio $N(\beta)/N(-\beta)$ for W1, V1 and Q1 band separately, where
$N(\beta)$ is the three-year observation number contributing to the pixels 
with latitude within $\beta\pm1.5^o$ 
in the northern hemisphere and $N(-\beta)$ is that in the south, 
the results are shown in Fig.~9. 
From Fig.~9 we see that there exists north-south asymmetry in WMAP sky survey: 
exposure increases
from south to north with maximum amplitude $\sim 2\%-4\%$. It is natural to explain the rms asymmetry
by the exposure asymmetry as $\sim 10\%-20\%$ decreasing of rms deviation 
is just expected by $\sim 2\%-4\%$ increasing of exposure, provided that the linear combination 
can not suppress the common and systematic north-south asymmetry in multi frequencies 
and the exposure dependent 
noise follows Eq.~\ref{sigma}, rms $\propto 1/\sqrt{(N_i)}$.
 
An apparent asymmetry in the distribution of fluctuation power in two opposing hemispheres
with a remarkable absence of power in the vicinity of the northern ecliptic pole is observed
from the WMAP1 Q, V, and W band sky maps \cite{eri04b, han04}.
By analyzing the issue of power asymmetry with the WMAP3 ILC map and a  model 
of an isotropic CMB sky modulated by a dipole field, Eriksen et al. (2007)  find that 
the modulation amplitude is $11.4\%$ and that the results on hemispherical power asymmetry 
are not sensitive to data set or sky cut.
All of these features can be explained naturally by the effect of exposure dependent noise
as we discuss above in this section.  

\clearpage
\section{Foreground contamination} 
 Besides the statistical noise, another kind of error in recovered WMAP CMB maps 
is systematic distortion.
From Eq.~\ref{mm-l} we can see that in an iteration the temperature estimation at a sky pixel $i$  
is evaluated based on temperatures of many pixels on a circle in the sky sphere 
 $\theta_{beam}$ (beam separation) apart from the pixel $i$, including $N_i^-$ pixels 
pointed by minus-horn
when plus-horn pointing to $i$ and $N_i^+$ pixels pointed by plus-horn
when minus-horn pointing to $i$. The sky temperature $t_{k^+}$ 
observed by plus-horn and $t_{k^-}$ by minus-horn are differently placed
in the right side of the iterative formula Eq.~\ref{mm-l}.
The iterative solution $\hat{t}_i$ could be deviated from 
the true temperature $t_i$ due to inhomogeneity of temperature sky 
and unbalance between two sky side beam measurements.

\subsection{Map distortion by a hot source}
The foreground induced systematic effect on a recovered map is not
only limited in the region containing foreground sources, but spreated
 over the whole sky. 
A hot foreground source pointed by the side beam A of a radiometer can 
distort the recovered temperatures of sky pixels pointed by the side beam B
with separation angle $\theta_{beam}$ to the source
through map-making iterations. From Eq.~\ref{mm-l} (or Eq.~\ref{mm-w}) the first 
iterative solution of temperature of a sky pixel $i$ can be expressed as
 \begin{equation} 
\label{t1}
t^{(1)}_i=t_i-\<t-t^{(0)}\>_{ring} \end{equation}   
where $\<~ \>_{ring}$ denotes averaging on the scan ring 
with separation angle $\theta_{beam}$ to the pixel $i$.
For zero initials $t^{(0)}=0$
\begin{equation}
\label{t10}
t^{(1)}_i=t_i-\<t\>_{ring}~, \end{equation}  
a hot source contained on the scan ring will let  $\<t\>_{ring}>0$ and 
the recovered temperature $t_i^{(1)}<0$, indicating that a hot foreground 
source might systematically make the recovered  temperatures on its scan ring lower.    

A one-dimensional simulation is made to show such effect.
A true temperature ``sky'' containing 500 pixels is produced 
as a white noise series of zero mean and 0.2 mK standard deviation 
with a hot source in the pixel interval 240-260, as shown in the upper panel of Fig.~10.
To produce differential data, we simulate a scan process of a differential 
radiometer with instrument noise $\sigma_0=4.0$ mK and beam separation  
 $\theta_{beam}=100$ pixel.  
From $i=1$ to 500, for each pixel $i$ we produce 2500 temperature differences 
$t_i-t_j+n$ where $j=i+100$ for $i\le400$ or otherwise $j=i-400$, 
$n$ is sampled from the normal distribution with zero mean and standard 
deviation $\sigma_0$.
We use Eq.~\ref{mm-w} to reconstruct temperature map from the differential data.
For pixel $k=350$ the recovered temperatures  from first fifteen iterations,
$t_k^{(1)}, t_k^{(2)}, \cdots, t_k^{(15)}$ 
are shown in Fig.~11, where we see 
that the recovered temperature of first iteration for the pixel
 $\theta_{beam}$ away from the hot pixel really drops down dramatically
as expected by Eq.~\ref{t1} and after a few iterations converged to a stable
value which still significantly lower than its real temperature.   
The lower panel of Fig.~10 shows the recovered temperature map 
after 50 iterations.
There exist two significant cold regions in the recovered map 
with pixels $340-360$ and $140-160$, 
each have a distance of $\theta_{beam}$ to the hot source interval
and their temperature variation negatively related to that of the hot source. 
Two more regions distorted in the similar way  
can be found in the recovered map, each 
placed 100 pixel away from the above two cold regions respectively,
pixel intervals $440-460$ and $40-60$. 
The above simulation result indicates that a hot source can distort 
temperatures in the map recovered from 
differential data at pixels away from the source with distance of 
the beam separation and such distortion can propagate to more far away pixels.       
  
\begin{figure}[htb]
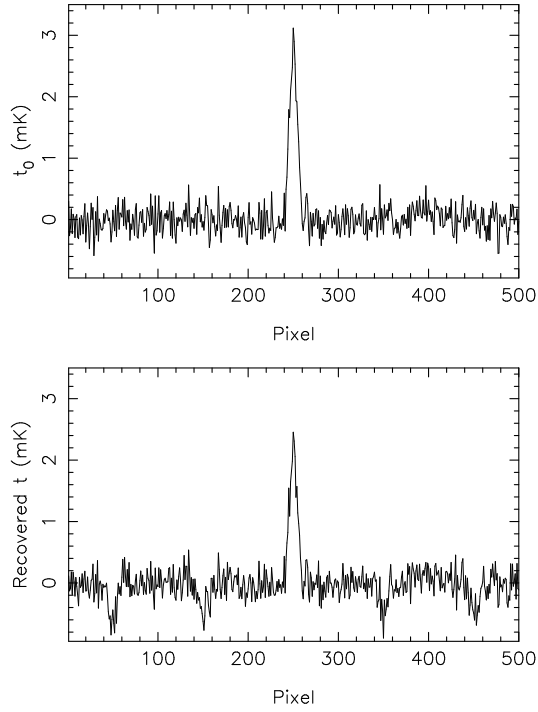

\label{f10}
   \begin{center}
    \psfig{figure=f10a.ps,width=45mm,height=70mm,angle=270}\\
\vspace{3mm}\psfig{figure=f10b.ps,width=45mm,height=70mm,angle=270}
   \caption{True and recovered temperatures. {\sl Upper panel}:
True temperatures. 
{\sl Lower panel}: Recovered temperatures. 
}
   \end{center}
\end{figure}

\begin{figure}[htb]
\label{f11}
   \begin{center}
    \psfig{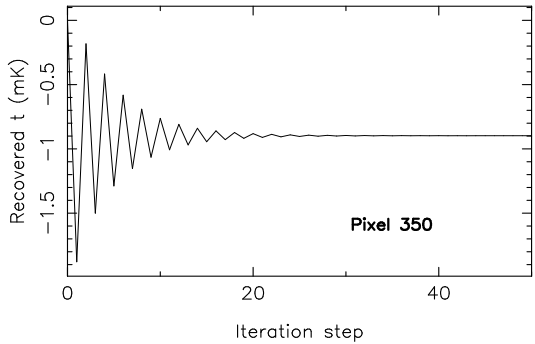}
   \caption{Recovered temperature at pixel 350 vs. step number of iteration by 
map-making Eq.~\ref{mm-w}. The one-dimensional true temperature map,
shown in the upper panel of Fig.~10,  has a hot source with
a peak at pixel 250, and the differential data is simulated 
for a radiometer with beam separation of 100 pixel. 
}
   \end{center}
\end{figure}
    
\subsection{$141\degree$ rings in WMAP maps}
The beam separation angle $\theta_{beam}$ of WMAP radiometers  is $141\degree$. 
 We predict from the above simulation that there should exist strongest negative
correlation in WMAP temperature maps between temperatures of two sky pixels 
separated $141\degree$ each other. This prediction is confirmed 
by analyzing temperature maps
in Q, V and W bands released by WMAP team. 
For the WMAP3 Q-, V- and W-band maps with HEALPix resolution parameter 
$N_{side}=128$ \cite{gor05}, we calculate cross-correlation coefficients 
$C_{t_a\<T_b\>}(\theta)$ between $t_a$ and $\<t_b\>$ for different
separation angle $\theta$, 
where $t_a$ denotes the temperature of a sky pixel $a$ and 
$\<t_b\>$ the average temperature of the ring with separation angle $\theta$ 
to the pixel $a$, the obtained distributions are shown in Fig.~12.

\begin{figure}[tb]
\label{f12}
   \begin{center}
\vspace{-4mm}\psfig{figure=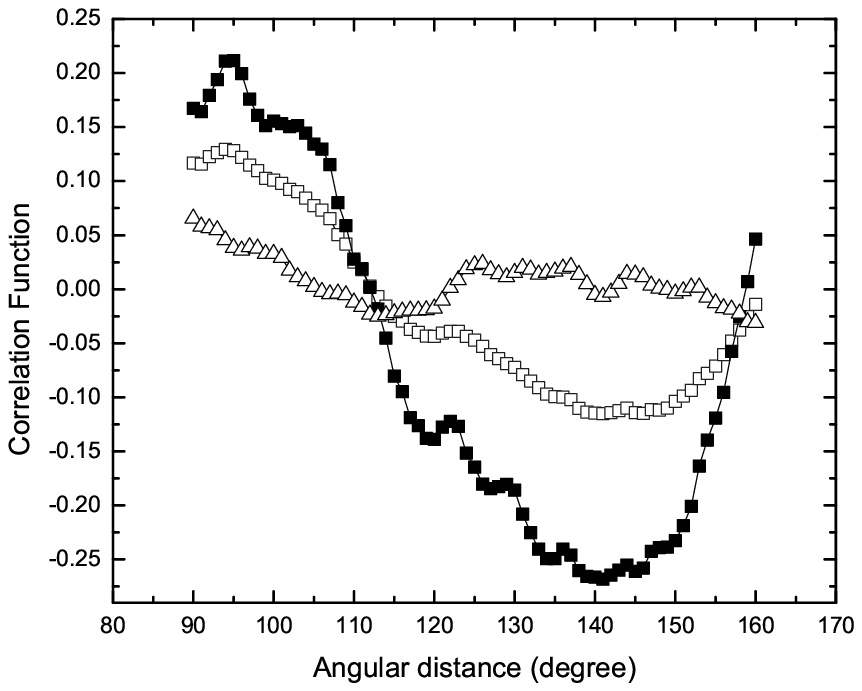,width=65mm,height=42mm,angle=0}\\
\vspace{-4mm}
    \psfig{figure=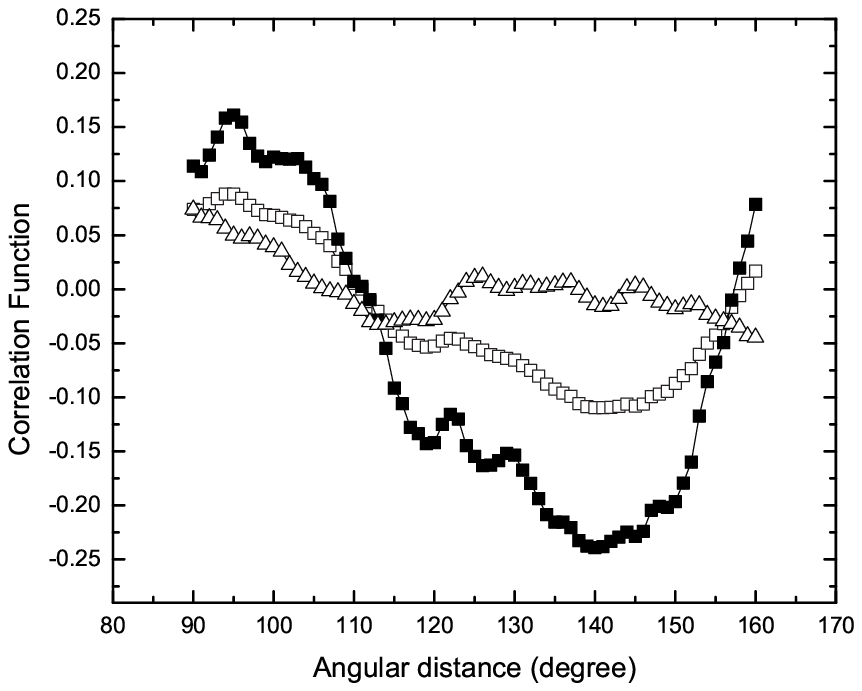,width=65mm,height=42mm,angle=0}\\
\vspace{-4mm}
    \psfig{figure=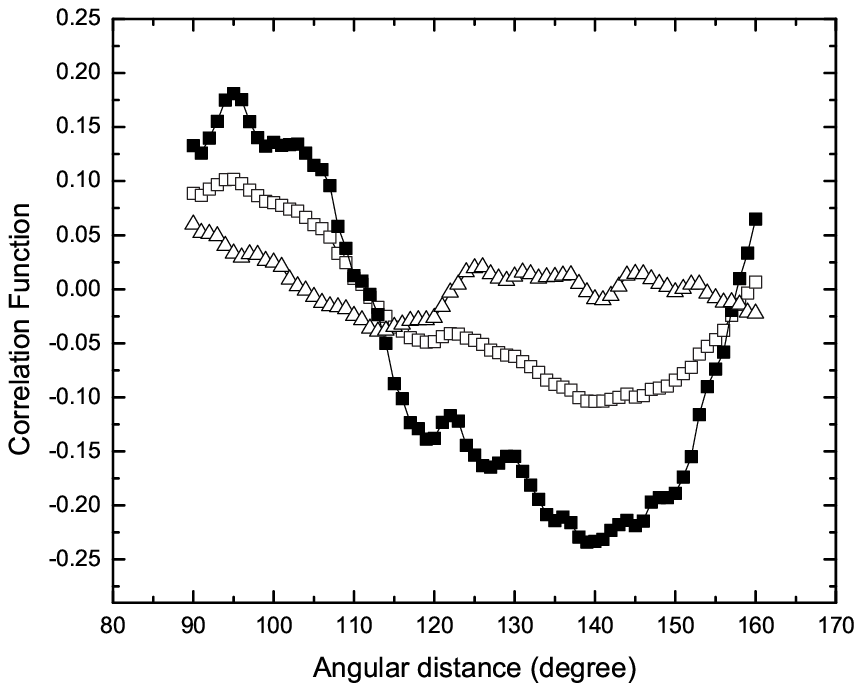,width=65mm,height=42mm,angle=0}
\vspace{-5mm}   \caption{Correlation coefficient $C_{t_a\<t_b\>}(\theta)$ between 
temperatures $t_a$ and $\<t_b\>$ vs. separation 
angle $\theta$. $t_a$ denotes the temperature of sky 
pixel $a$, 
$\<t_b\>$ the average temperature of the ring with separation angle $\theta$ to $a$.
{\it Filled square}: pixel $a$ only within the foreground mask Kp12.
{\it Square}: for whole sky. 
{\it Triangle}: only for the sky region out of the mask Kp0.
{\sl Top panel}: WMAP3 Q-band.
{\sl Middle panel}: WMAP3 V-band.
{\sl Bottom panel}: WMAP3 W-band.  } 
   \end{center}
\end{figure}  

From Fig.~12 we find that the strongest negative correlation appear
 indeed around $141\degree$ separation in the correlation distribution
for each band. 
If pixel $a$ is limited within the 
sky region of the foreground mask Kp12 \cite{ben03b} which consists mainly of hot pixels,
 around $141\degree$ separation show more strong negative correlation 
just as expected from assuming that 
hot foreground emission induce the correlation. 

The program synfast in HEALPix software package (available at
 http://healpix.jpl.nasa.gov)
can create temperature maps computed as realizations 
of random Gaussian fields on a sphere characterized by the user provided 
spherical harmonic coefficients of a angular power spectrum.
To test the significance of the correlation $C(141\degree)=-0.238$ 
in Q-band WMAP map, simulated temperature maps are created with 
the synfast program 
from the best fit $\Lambda$CDM model power spectrum with Q-band
beam function and noise property. For each simulated CMB map, 
we compute the correlation coefficient between temperatures of the pixels in Kp12 region 
and average temperatures on their $141\degree$ rings. From 1000 simulated maps 
we get $C^\prime(141\degree)=0.108\pm0.080$, indicating that the $141\degree$ 
negative correlation detected in WMAP Q-band map has a significance 
of $4.3\sigma$. Similarly, for V-band we get $C(141\degree)=-0.231$ 
with significance $4.4\sigma$ evaluated by 
$C^\prime(141\degree)=0.099\pm0.075$ from simulated CMB maps, 
and for W-band $C(141\degree)=-0.268$ with significance $5.0\sigma$ from
 $C^\prime(141\degree)=0.094\pm0.072$. 
 
\begin{figure}[p]
  \label{f13}
    \begin{center}
 \vspace{-10mm}\includegraphics[width=40mm,angle=90]{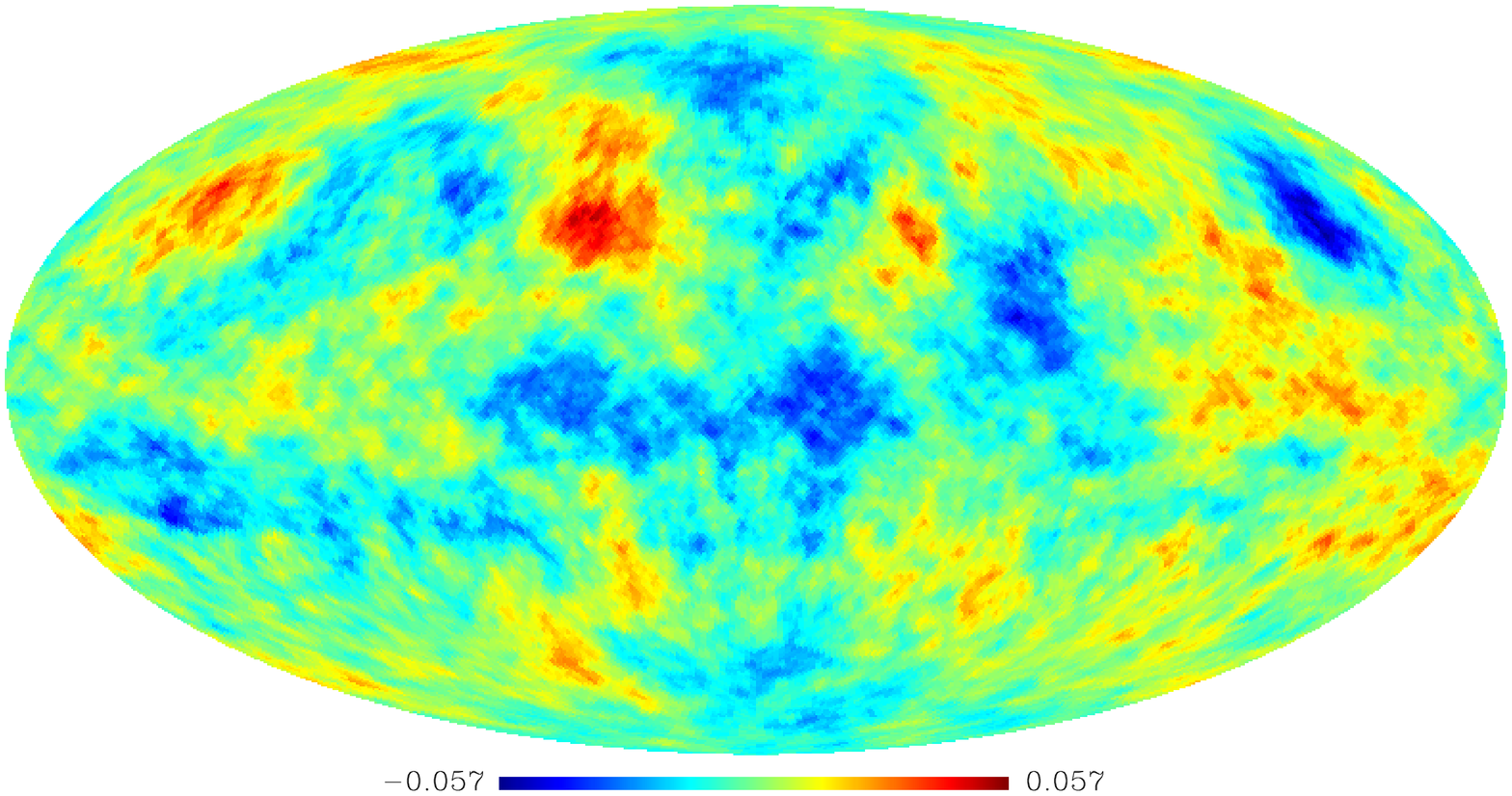}\\
 \vspace{-2mm}\includegraphics[width=40mm,angle=90]{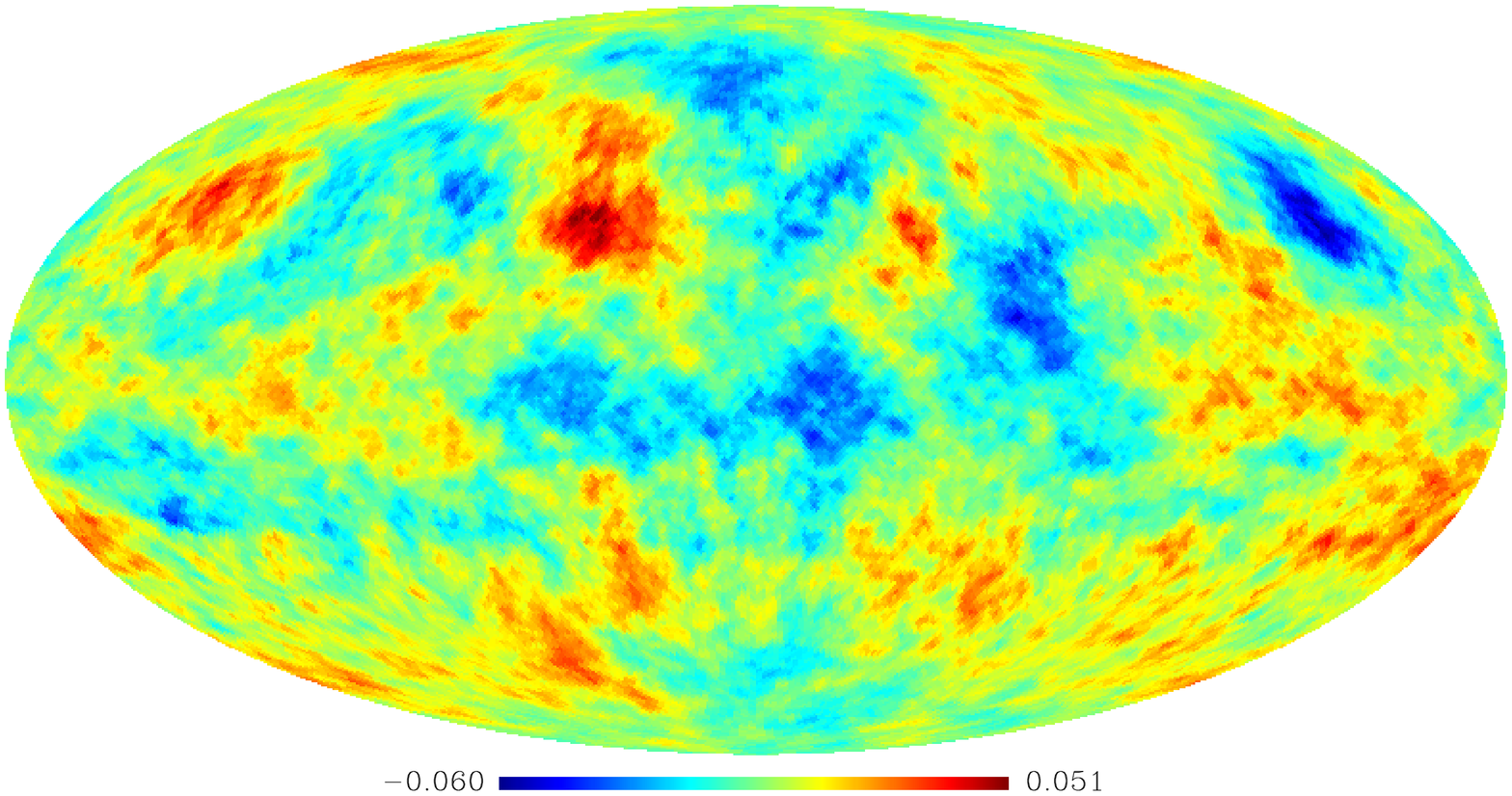}\\
 \vspace{-2mm}\includegraphics[width=40mm,angle=90]{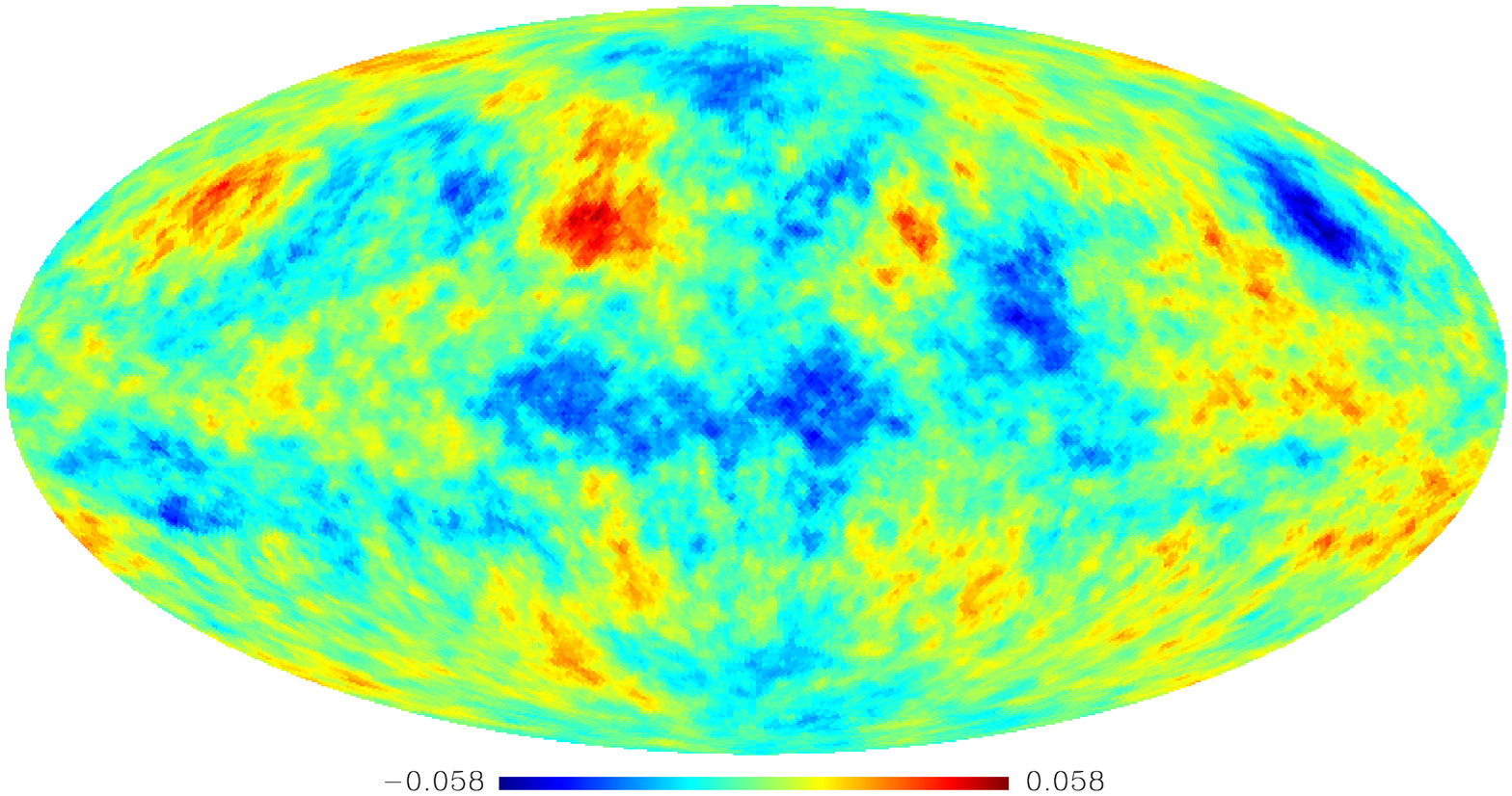}\\
 \vspace{-2mm}\includegraphics[width=40mm,angle=90]{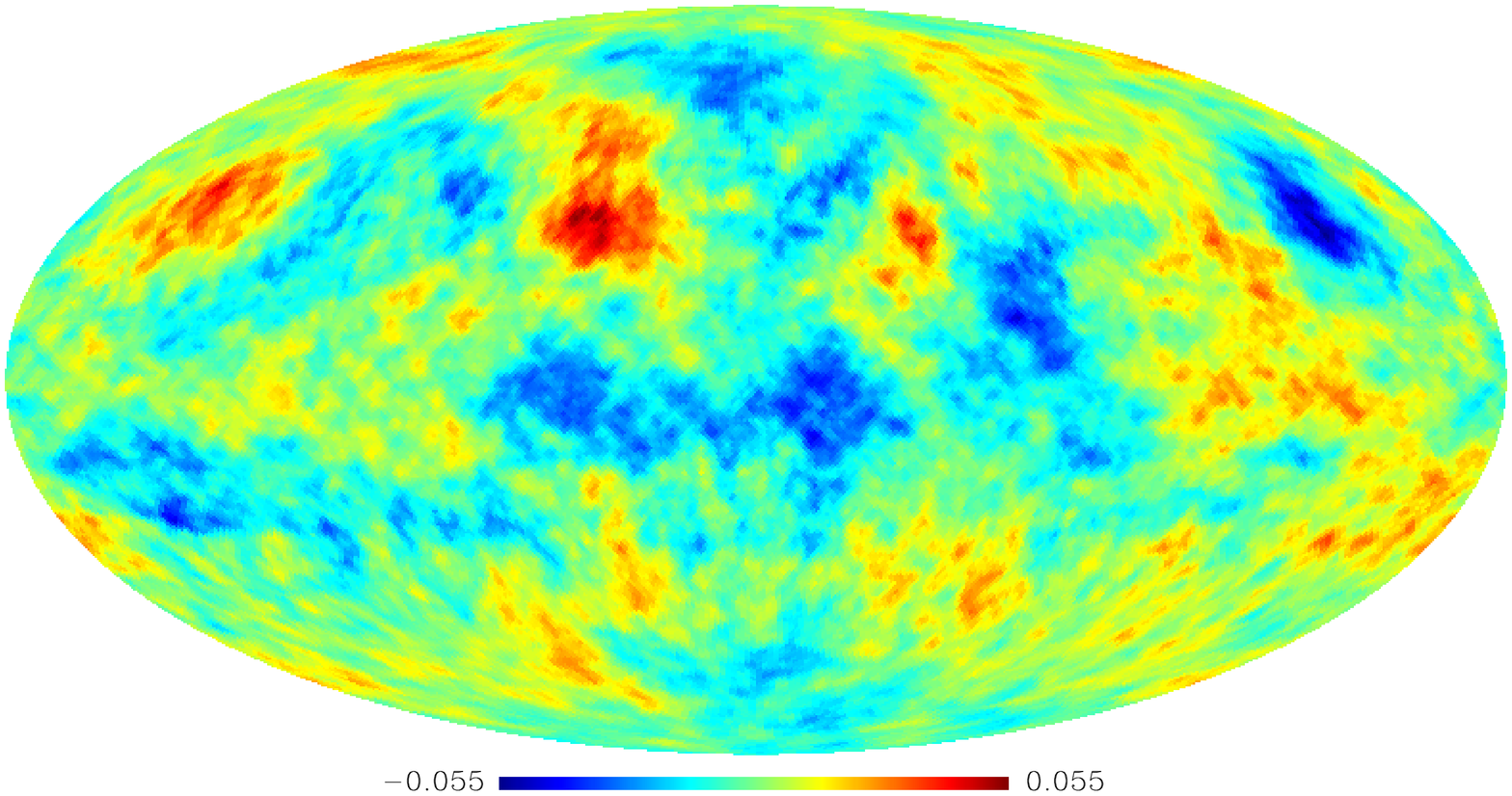}
 \caption{Average temperature maps of $141\degree$ ring in Galactic coordinates
from WMAP3 Q-band, V-band, W-band and ILC map (from top to bottom panel).
}
    \end{center}
\end{figure}
It is expected from our simulation shown in Fig.~10 that foreground hot sources can
produce cold rings of $141\degree$ separation in WMAP temperatures. 
To test this effect, for each sky pixel we calculate the average temperature 
of its $141\degree$ ring from the WMAP3 Q-band, V-band, W-band and ILC temperature
maps with $N_{side}=128$ and Kp0 mask \cite{ben03b} for foreground clean.
The result is shown in Fig.~13. There is a low temperature region 
near Galaxy center in Fig.~13 for each band 
indicating that most of the $141\degree$ rings corresponding to the galaxy center region  
are cold. This is consistent with our expectation.

To roughly estimate the magnitude of temperature distortion by foreground emission, 
we pick up 2000 hottest pixels in the 3-year WMAP Q-band map and then 
find out their $141\degree$ rings. The average temperature of all pixels on these rings 
and out of the Kp0 mask (cover about $14\%$ of the sky, see Fig.~14) is calculated to be 
$\<t\>=-0.0117$ mK. 
The correspondent value from 1000 simulated CMB maps is $\<t^\prime\>=3.4\times10^{-5}\pm0.006$ mK.
Similarly, for V-band we get $\<t\>=-0.0125$ mK and  $\<t^\prime\>=-4.7\times10^{-5}\pm0.006$ mK,
for W-band  $\<t\>=-0.0141$ mK and  $\<t^\prime\>=8.8\times10^{-5}\pm0.006$ mK.
Therefore, in foreground-cleaned WMAP maps, considerable foreground-induced deviations 
still exist with amplitude comparable to the fluctuation of CMB signal and 
over broad sky region.   

\begin{figure}[htb]
\label{f14}
    \begin{center}
    \includegraphics[width=48mm,angle=90]{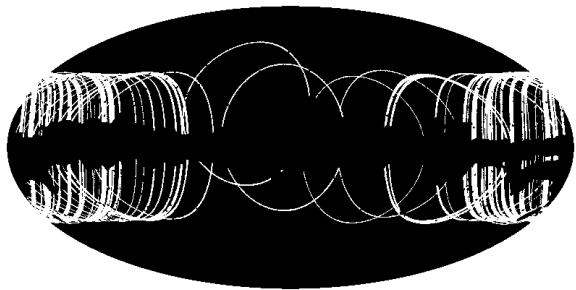}
   \caption{$141\degree$ rings of 2000 hottest pixels in
3-year Q-band WMAP map, after using the Kp0 mask, in Galactic coordinates.
}
    \end{center}
\end{figure}

The foreground-induced deviation in WMAP temperatures  should also distort the temperature 
angular power spectrum. We use simulated temperature maps created with 
the synfast program to roughly estimate the magnitude of foreground-induced distortion     
for the WMAP power spectrum on large-scale (low-$l$) region. 
For each simulated map, we subtract 0.01 mK from each temperature of pixels
on the $141\degree$ rings corresponding to the 2000 hottest pixels (shown in Fig.~14) 
to get a distorted map. We compute the power spectrum for simulated and 
distorted map respectively and find that, on average, the deviation between the original
and distorted spectrum densities at $l=2, 3$ is $\sim 10\%$,
indicating that the foreground-induced distortion on WMAP power spectrum
cannot be ignored for a precision cosmology study.
Strongest hot sources may produce cold spots by the foreground induced distortion
out of the foreground region.
Temperature distortions in a CMB map caused by foreground sources with different 
scales and observation dependent noise could distort the angular power spectrum
on wide range of angular scale.  

\subsection{Large non-Gaussian spots}
A large cold spots with a radius of $\sim 5\degree$ 
centered at $(l,b)=(209\degree, -57\degree)$ and the lowest 
temperature $\sim -0.15$ mK has been 
detected in WMAP1 and WMAP3  maps with the wavelet and other techniques 
\cite{vie04,cru05,cru06,cru07a,vie07},
and was explained as a cosmic texture, a remnant of symmetry-breaking 
at energies close to the Planck scale in the very early universe~\cite{cru07b}.
We find that the $141\degree$ rings of the pixels in the spot region
across through the hot Galactic plane with temperature up to $\sim 5$ mK,
therefore, the foreground induced systematic effect on recovered maps 
discussed in \S3.1 should be a more plausible explanation to the detected large cold spot 
than the texture hypothesis. The foreground contamination and other systematic 
effect may also produce other more
detected large spots in WMAP maps. 

\section{Discussion}
\subsection{Can the exposure induced anisotropy be corrected ?}
It has to be pointed out that Eq.~\ref{rms} can be used to modify the effect of
exposure dependent noise, like what we do for the latitude distribution 
of rms variation shown in Fig.~2, only for the case that the map rms 
fluctuation $\<{\hat{t}}^2\>$ itself is the directly analyzed quantity.
However, almost all analysis works are based on CMB maps of temperature.
We know from Figs.~2-5 that the released WMAP temperature maps  
contain considerable exposure dependent noise.
It is no way  from a recovered temperature map to produce a corrected map
in which
the instrument-induced and exposure-dependent noise can be eliminated. 
Therefore, the anisotropy noise should contribute to large scale anomalies 
detected in existent CMB maps.

The exposure inhomogeneity of WMAP comes from its scan strategy, which can't
be suppressed through accumulating observation time.   
To avoid the observation effect, it is needed to remake temperature maps
from a uniform differential data set obtained by giving up partial observation data
for pixels of high exposure. Comparing released WMAP maps and new maps from uniform
data will help us to judge their origin of detected large scale anomalies, 
e.g. the low $l$ power issues detected in WMAP data, 
the unexplained orientation of large-scale patterns of CMB maps in respect 
to the ecliptic frame, the north-south asymmetry of temperature fluctuation power 
etc., and to see if the observational effect can also influence the angular
power spectrum as well. 

\subsection{Avoiding the foreground-induced distortion}
\begin{figure}[p]
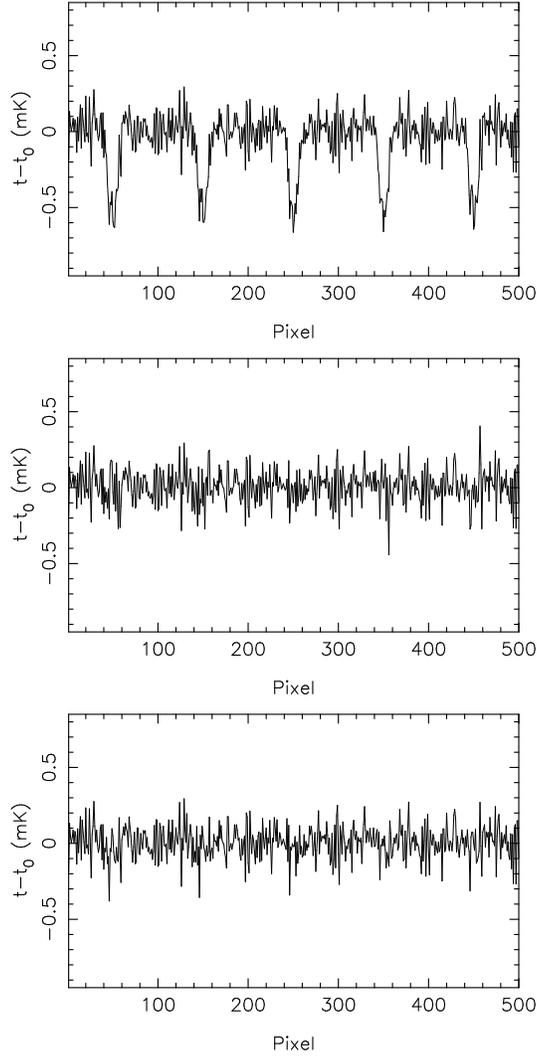

\label{f15}
   \vspace{-18mm}
   \begin{center}
    \psfig{figure=f15a.ps,width=45mm,height=70mm,angle=270}\\
\vspace{2mm}\psfig{figure=f15b.ps,width=45mm,height=70mm,angle=270}\\
\vspace{2mm}\psfig{figure=f15c.ps,width=45mm,height=70mm,angle=270}
   \caption{Differences between recovered $t$ and true temperature $t_0$.
The one-dimensional temperature distribution $t_0$ consists of white 
noise and hot source between 240 - 260 pixel (mask region), as shown 
in the upper panel of Fig.~10. Differential data is obtained by simulation
of one-dimensional scan for $t_0$ by a differential radiometer 
with beam-separation of 100 pixel. Recovered $t$ are calculated with map-making 
Eq.~\ref{mm-w} after 50 iterations. 
{\sl Upper panel}: $t$ recovered from all differential data (the lower panel of Fig.~10).  
{\sl Middle panel}: $t$ recovered by iterations from all data for mask region 
and with excluding differences that contain hot source temperature for pixels out of mask. 
{\sl Bottom panel}: $t$ recovered by iterations with initials estimated 
from the differential data by Eq.~\ref{tj}. }
   \end{center}
\end{figure}

The distortion by hot foreground sources on their $141\degree$ rings
in a WMAP map can not be removed with a foreground mask on the recovered temperature map.   
What's needed is to use the mask on the original differential data before map-making
to avoid the foreground-induced error in the recovered map.
The top panel of Fig.~15 shows the difference between the recovered and
true temperature distributions (shown in the lower and upper panel of Fig.~10 respectively),
where the distortion structure caused by the hot source on pixel 240 - 260 and 
beam separation of 100 pixel is clearly shown. We redo the temperature reconstruction
with excluding the temperature differences that contain a 
beam side pointing to a pixel between 240 - 260 (``mask region'') 
during iterations for the pixels out of mask, the result
is shown in the middle panel of Fig.~14, where the distortion structures
are really suppressed. 

A weakness of using mask in map-making process is decreasing the number of
useful differential data. Another approach to avoid the distortion in recovered map
by foreground emission is to properly set the initials of iteration for 
the foreground region. From Eq.~\ref{t1} we see that the temperature deviation
of the first iterative solution, $t_i^{(1)}-t_i$, will be suppressed if the 
temperature initials at pixels of hot source are set to be close to their true values
to let $\<t-t^{(0)}\>_{ring}\approx~0$.     
The initial $t_i^{(0)}$ of pixel $i$ can be taken as
\begin{equation}
\label{tj}
  t_i^{(0)}=\frac{1}{N_i^\prime}(\sum_{k^+=i}d_k-\sum_{k^-=i}d_k)~. 
\end{equation}
Where $\sum_{k^+=i}$ means summing over the observations while the pixel 
$i$ is observed by the plus-horn and the pixel pointed by the minus-horn 
is out of mask, $\sum_{k^-=i}$ means summing over the 
observations while the pixel $i$ is observed by the minus-horn and
the pixel pointed by the plus-horn is out of mask, 
$N_i^\prime$ is the total number of used observations. 
For the simulated differential data from the true temperatures shown in the upper 
panel of Fig.~10,  we make 50
iterations with Eq.~\ref{mm-w} starting from initials calculated by Eq.~\ref{tj}, 
the distortion structures are satisfactory suppressed 
in the resultant solution, as shown in the bottom panel of Fig.~15.
 
\subsection{Remaking WMAP maps}
We demonstrate in this paper that for existent CMB maps 
both the observation dependent noise and 
systematic error induced by foreground emission can not be 
neglected and both can produce large-scale anomalies 
and distort the angular power spectrum.
These errors can not be completely excluded by performing noise suppressing or 
using foreground mask on temperature maps.
We suggest to remake temperature maps from the original WMAP time-order-data
by a modified algorithm with applying foreground mask in map-making 
to exclude mask pixels from use
in iterations for CMB dominated region (or properly set 
temperature initials before iteration),
and/or keeping
used differential data uniform by giving up partial observation data
for pixels of high exposure. 
New maps from modified map-making algorithm
will help us to judge the origin of large scale anomalies detected in released 
WMAP maps, e.g. the low $l$ power issues, 
the unexplained orientation of large-scale patterns in respect 
to the ecliptic frame, the north-south asymmetry of temperature fluctuation power 
and the large non-Gaussian spots, 
and to see to what extent the statistical and systematical errors 
influence the angular power spectrum and the derived cosmological parameters.
Believable conclusions on CMB anisotropy anomalies and precise temperature 
angular power spectrum from differential measurement
should be based on temperature maps with homogeneous sky exposure
and should avoid foreground-induced distortion during map-making.
 
This study is supported by the National Natural 
Science Foundation of China and the CAS project KJCX2-YW-T03.  
The data analysis in this work made use of the 
WMAP data archive and the HEALPIX software package.

\end{document}